# Multi-cell RF Injectors Driven by Thermoionic Cathodes

N. Piovella, L. Serafini, University of Milan and INFN, Milan, Italy
M. Ferrario, INFN-LNF, Frascati, Italy

*Abstract*

We present the anticipated performances of a multi-cell RF injector operated either with a thermoionic cathode or with a gated photo-cathode. By means of a magnetic compression and a Coherent Spontaneous Emission seeded FEL radiating at 100 to 200 µm, the system produces a beam fully bunched with a 200 A peak current at 2 mm·mrad rms emittance. The electron bunch is a few FEL buckets long and phase-locked to the RF so to make this device an ideal injector for second generation plasma acceleration experiments.

## 1 LASERLESS MULTI-CELL RF INJECTORS

The basic requirements concerning the electron beam to be injected into a plasma acceleration experiment of second generation can be summarized as follows [1]:

i) injection must be performed at about 10 MeV of beam energy to preserve beam quality (bunching and emittance) and the beam must be pre-bunched on the scale of the plasma wavelength, which ranges between 100 to 300 µm, depending on the type of acceleration scheme to be applied (wake field or beat wave). This implies the need to generate a train of bunchlets 50-100 fs long, each one carrying $10^8$ electrons, spaced 0.3 to 1 ps far apart.

ii) the bunching structure in the electron pulse must be phase locked to the laser driving the plasma wave in order to guarantee a definite synchronous acceleration phase in the plasma wave. This implies that the phase jitter must be smaller than a fraction of the plasma wave period, *i.e.* no larger than 50-100 fs.

iii) the rms normalized emittance must be smaller than a few mm·mrad to guarantee a matching of the beam at injection into the plasma wave, whose transverse size is about 50-100 µm.

RF Laser-driven Photo-Injectors are unfortunately unable, to the present state of the art, to comply with requirement i) due to the tight phase jitter requested [1], though the minimum bunch length achievable in single bunch mode can be as low as 100 fs [2]. On the other hand, even Free Electron Lasers, who are excellent candidates for the generation of bunching in the sub-millimiter range, suffer from non-uniformity of the bunching over the electron pulse because of slippage effects (especially when they are operated in the single pass high gain regime in steady state) [3].

In order to join the benefits of both devices (high brightness capability of RF guns and bunching action of FELs) we propose to use a multi-cell RF injector driven either by a thermoionic cathode or by a gated photocathode, coupled to a magnetic compressor and a FEL which is seeded by the Coherent Spontaneous Emission (CSE) driven by the initial bunching associated to the sharp current profile in the electron pulse head. The electron emission at a thermoionic cathode takes place all over the RF period: this is possible also by illuminating a photocathode with a laser pulse longer than the RF period (which does not need, in this case, to be phase-locked to the RF as in the usual operation of laser driven photoinjectors). In both cases the phase jitter of the electron bunch at the gun exit is no longer dependent on the launching phase jitter at the cathode, which is the main source of phase fluctuation in the beam.

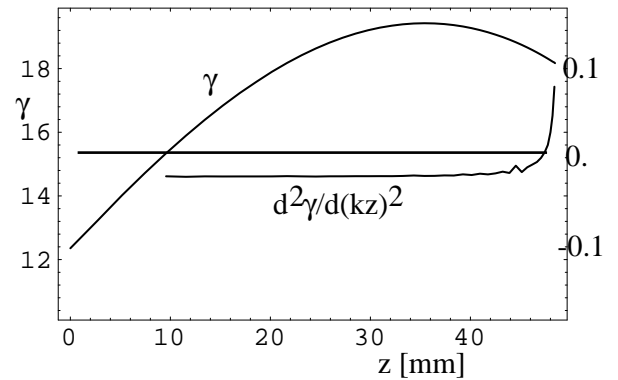

Figure 1: Energy as a function of position in a quarter of RF period long bunch extracted from a multi-cell RF gun. Its second derivative is also plotted.

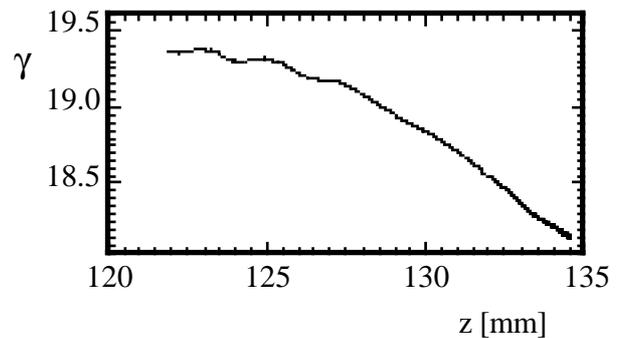

Figure 2: Longitudinal phase space density distribution at the gun exit before magnetic compression: only the 20 RF deg at the bunch head are plotted.

When electrons are emitted from the cathode surface (located in the first half cell of a RF gun cavity) all along the RF period, only those who are emitted between the null-point of the RF electric field (the 0° RF phase) and the peak field phase (90° RF) are actually trapped by the accelerating field and accelerated up to the gun exit. The exit beam has in this case a huge energy spread, as shown in fig.1, which reports a 1D (no space charge) calculation of the energy-phase relationship at the exit of a 6+1/2 cell gun working at 1.3 GHz with 25 MV/m peak RF field at the cathode.

A 2D C.I.C. self-consistent simulation performed with the code ITACA shows a similar behavior in the bunch head, with some energy modulations induced by the longitudinal space charge at the emission from the cathode surface, as shown in fig.2.

The interesting property of such a beam is the highly linear behavior of the energy-phase relation in the bunch head (the first 20° RF, *i.e.* between z=38 and z=49 mm), as indicated by the second derivative null-point at z=48 mm. A magnetic compression applied downstream, *e.g.* by means of a chicane compressor (or a wiggler) will give rise to a relevant current enhancement in that part of the beam, while it will dilute the tail of the beam which has an anti-correlated energy-phase correlation [2].

We simulated the magnetic compression occurring in a wiggler 5 period long, with a 11 cm wiggler period and 9 kG applied peak field, obtaining the phase space distribution and current profile reported in fig.3, which shows the output of a 2D ITACA run through the wiggler (a similar simulation has been repeated with the code Parmela to check 3D effects, obtaining similar results).

The peak current has increased from a 20 A, almost flat distribution along the bunch (as in fig.2) up to 150 A, with a sharp current peak in the region of the bunch head. This naturally suggests the presence of a non negligible bunching on the scale of sub-millimiter range in this part of the bunch, so to conceive the application of a FEL interaction to enhance further this bunching.

To this purpose we modelled, using the new code SPECTRE developed to study CSE processes [4], the FEL interaction occuring when the beam is injected into a 2 m undulator, whose period is 2 cm long, with wiggler parameter $a_W$=1.5. The code is able to model time dependent FEL processes in 1D with a multi-frequency analysis.

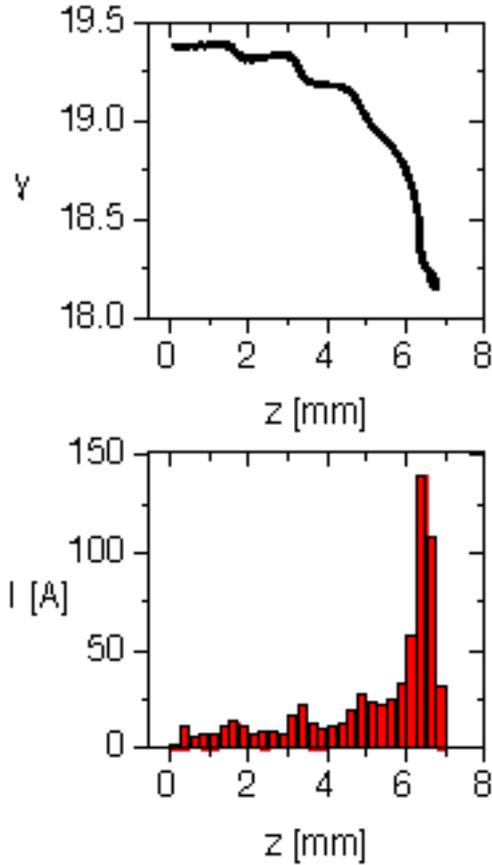

Figure 3: Longitudinal phase space density distribution after magnetic compression (upper diagram) and corresponding current profile (lower diagram).

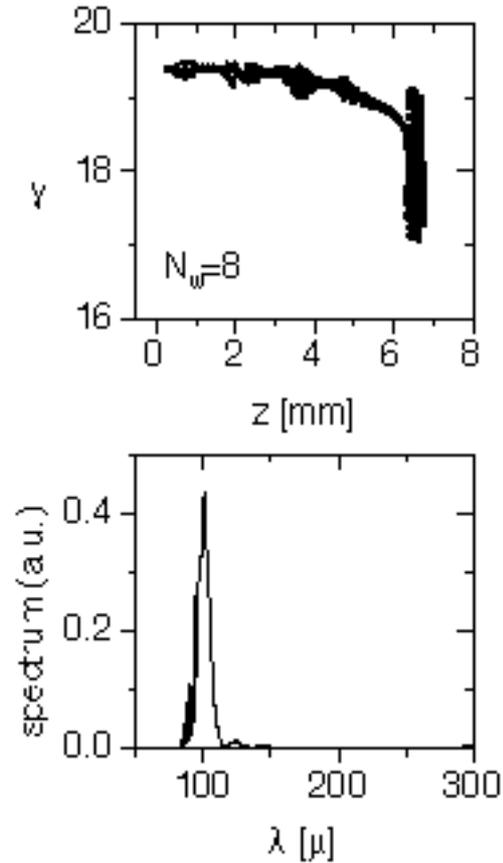

Figure 4: Longitudinal phase space density distribution after 8 periods of the FEL undulator (upper diagram) and corresponding FEL radiation spectrum (lower diagram).

The beam distribution after 8 undulator periods is plotted in fig.4 (upper diagram). The FEL interaction is clearly started by a strong CSE effect due to the current distribution in the bunch head: the few FEL buckets there located are already strongly radiating at 100 μm wavelength, as shown by the FEL radiation spectrum in the lower diagram.

While the FEL interaction proceeds, these buckets tend to split from the rest of the beam, which is still in a FEL lethargy regime, and they decrease in energy due to the radiation emission process. Figure 5 clearly shows the onset of such a process, together with the corresponding frequency sweep of the FEL radiation, moving toward longer wavelengths as these FEL buckets loose energy.

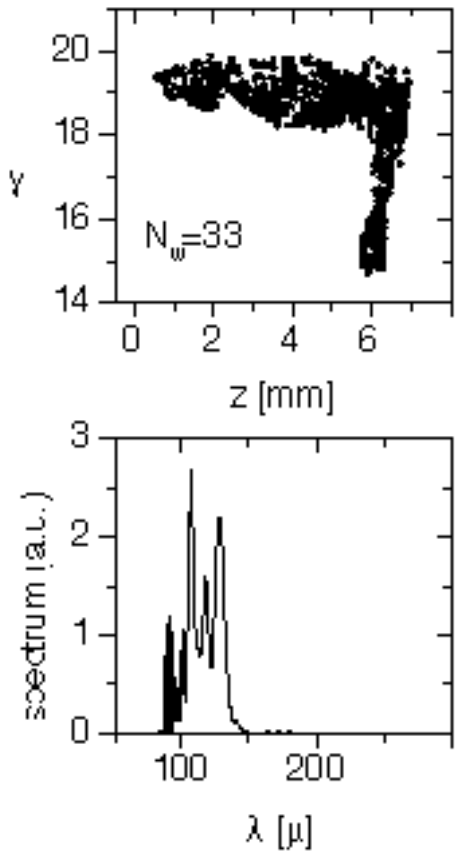

Figure 5: Longitudinal phase space density distribution after 33 periods of the FEL undulator (upper diagram) and corresponding FEL radiation spectrum (lower diagram).

Eventually, after 1.5 m of undulator length (see figure 6), the bunch has split into two separate parts: the former bunch head, which is now slipping backward and exiting through the beam tail, and the beam core (the former bunch tail) which is now radiating at the original 100 μm resonant FEL wavelength.

The very appealing property of the split bunch is the uniformity of the bunching on the scale of 250 μm (5 buckets are present) ad the possibility to easily separate it from the rest of the beam, due to the relevant energy difference.

## CONCLUSIONS

Since the absolute phase of these buckets is determined only by the RF phase in the gun (no phase locked laser pulse has been used to produce the beam!), the only source of phase jitter is given by the FEL interaction, which is anyway stabilized by the seed produced by CSE. The final phase jitter at injection into the plasma wave is anticipated to be small, though further analysis on this issue has to be carried out in the future.

Transverse dynamics in the FEL will also have to be analyzed, in order to assess the possible emittance dilution occurring in the undulator: though the requirements of the plasma acceleration experiments [1] are typically for emittances not larger than 5 mm·mrad, and the beam emittance in the bunch head at the exit of the compressor (fig.3) is about 3 mm·mrad, this issue has certainly to be checked.

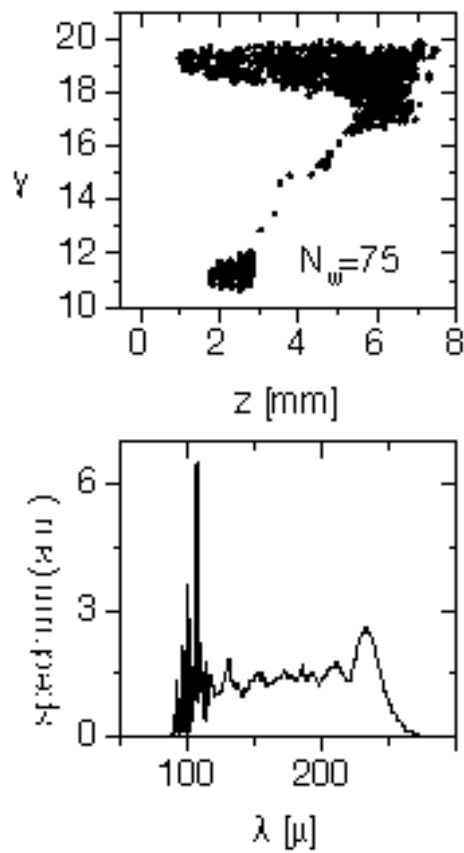

Figure 6: Longitudinal phase space density distribution after 75 periods (1.5 m) of the FEL undulator (upper diagram) and corresponding FEL radiation spectrum (lower diagram).

## REFERENCES

[1] C. Clayton and L.Serafini, IEEE Trans. on Plasma Science **24** (1996) 400.
[2] L.Serafini, IEEE Trans. on Plasma Science **24** (1996) 421.
[3] C. Pellegrini et al., IEEE Trans. on Plasma Science **24** (1996) 537.
[4] N. Piovella, AIP CP **413** (1997) 95.